\documentclass[twocolumn,aps,pra,floatfix,amsmath,amssymb,showpacs,nofootinbib]{revtex4} 
\usepackage{graphicx}
\usepackage{hyperref}

\newcommand{\ket}[1]{\ensuremath{|{#1\rangle}}} 

\newcommand{\braket}[2]{\ensuremath{{\langle #1}|{#2 \rangle}}}
\newcommand{\ketbra}[2]{\ensuremath{|{#1 \rangle}{\langle #2}|}}
\newcommand{\op}[1]{\hat{#1}}
\newcommand{\e}{\text{e}}
\providecommand{\abs}[1]{\left\lvert#1\right\rvert}

\begin{document}

\title{Classicality, the ensemble interpretation, and decoherence:\\
  Resolving the Hyperion dispute}

\author{Maximilian Schlosshauer}
\email{m.schlosshauer@unimelb.edu.au}
\affiliation{School of Physics, The University of Melbourne,
  Melbourne, Victoria 3010, Australia}
  
\date{\today}

\begin{abstract}
  We analyze seemingly contradictory claims in the literature about
  the role played by decoherence in ensuring classical behavior for
  the chaotically tumbling satellite Hyperion. We show that the
  controversy is resolved once the very different assumptions
  underlying these claims are recognized. In doing so, we emphasize
  the distinct notions of the problem of classicality in the ensemble
  interpretation of quantum mechanics and in decoherence-based
  approaches that are aimed at addressing the measurement problem.
\end{abstract}

\pacs{03.65.Sq, 03.65.Yz, 05.45.Mt}

\maketitle

\section{Introduction}

The question of how the classical world of our experience can be
explained from within quantum mechanics continues to fuel a lively
debate. At the heart of this problem of the quantum-to-classical
transition is the superposition principle, which is formally grounded
in the linearity of the Hilbert space. Since quantum states are
represented by vectors in a Hilbert space, we may form linear
combinations of these vectors. The superposition principle then states
that such linear combinations correspond again to a new quantum
state. Superpositions cannot be interpreted as classical ensembles of
their components states. Instead, the phenomenon of interference shows
that all components in the superposition must be understood as, in
some sense, simultaneously present.

A particularly counterintuitive instance are systems described by a
superposition of macroscopically distinguishable positions. One way
such superpositions may dynamically arise is via a von Neumann
measurement \cite{Neumann:1932:gq} of a microscopic system prepared in
a certain superposition state. Unitary evolution applied to the
composite system--apparatus combination may then lead to an entangled
superposition state whose components refer to the pointer of the
apparatus being located at distinct positions on the dial. Another,
and rather different, example is the coherent spreading of initially
localized wave packets. Suppose a free particle is at time $t=0$
described by a wave packet of the form
\begin{equation}
\psi(x,t=0) = \left( \frac{1}{\sqrt{\pi}\sigma} \right)^{1/2}
\exp\left[ - \frac{x^2}{2\sigma^2} \right],
\end{equation}
then the position probability density $\abs{\psi(x,t)}^2$ at $t > 0$
is given by
\begin{equation}
\abs{\psi(x,t)}^2 = \frac{1}{\sqrt{\pi}\sigma(t)} 
 \exp\left[ - \frac{x^2}{\sigma^2(t)}
  \right],
\end{equation}
where the width $\sigma(t)$ of the wave packet grows as 
\begin{equation}
\label{eq:kjdshvn2}
\sigma(t) = \sigma \sqrt{1 + \hbar^2 t^2/(m^2\sigma^4)},
\end{equation}
with $m$ denoting the mass of the particle. For microscopic particles
the spreading of the wave packet occurs on extremely short
timescales. For example, for an electron ($m \approx 10^{-30}$~kg) and
$\sigma(t=0) = 1$~\AA, we obtain $\sigma(t=1~\text{s}) = 10^{16}\sigma
= 10^6$~m. This coherent spreading was the core obstacle encountered
by Schr\"odinger when he initially tried---inspired by ideas laid out
in de Broglie's Ph.D.\ thesis of 1924 \cite{DeBroglie:1924:rx}---to
directly identify narrow wave packets with particles
\cite{Schrodinger:1926:pz}. The concept of particles is virtually
synonymous with localizability, a feature evidently not described by a
wave packet that may rapidly and coherently disperse over
macroscopic distances.

On the other hand, it seemed possible to uphold a peaceful
correspondence between wave packets and well-localized objects in at
least two situations. The first example, studied by Schr\"odinger in
1926 \cite{Schrodinger:1926:pz}, is the special case of coherent
states for the quantum harmonic oscillator, where the wave packet
remains narrow at all times $t >0$ and where its peak oscillates back
and forth similar to a classical point mass. The second example, which
will be of most interest for the purpose of this paper, is represented
by macroscopic systems for which the spreading described by
\eqref{eq:kjdshvn2} is \emph{typically} very slow.

However, the rate of spreading may be drastically enhanced in chaotic
systems irrespective of their size, and thus the problem of coherent
spreading of the wave packet over large regions in space may reappear
even at the level of macroscopic systems. This situation was studied
by Zurek \cite{Zurek:1998:om} using the example of Hyperion, a
chaotically tumbling moon of Saturn. Chaotic systems exhibit an
exponential sensitivity to the initial phase-space parameters. Wave
packets may diverge or become squeezed in the position or momentum
direction at an exponential rate given by the Lyapunov exponent
$\lambda$. Suppose the initial spread in momentum is $\Delta p_0$ and
exponential squeezing in this direction occurs, $\Delta p(t)= \Delta
p_0 \e^{-\lambda t}$. In the quantum setting, it follows from the
uncertainty principle that the initial spread in position must be at
least on the order of $\hbar/\Delta p_0$, and the required
conservation of the phase-space volume and unitary dynamics then imply
that the wave packet undergoes coherent spreading according to $\Delta
x(t) \sim (\hbar/ \Delta p_0) \e^{\lambda t}$. Zurek estimated that
within $\sim 20$~years the quantum state of Hyperion would evolve into
a highly nonlocal coherent superposition of macroscopically
distinguishable orientations of the satellite's major axes, thus
setting up a measurement-free version of Schr\"odinger's cat. In the
absence of decoherence, it would always be possible to choose an
appropriate observable that would confirm the existence of this
superposition, either through a direct projective measurement onto the
superposition state itself or by means of an interference measurement
in the component basis.

Using the standard model for quantum Brownian motion, Zurek then
showed that decoherence rapidly suppresses such coherent spreading,
locally degrading the superposition into an apparent (improper)
ensemble of narrow position-space wave packets. The superposition
initially confined to the satellite is rapidly dynamically
dislocalized into the composite satellite--environment system via
environmental entanglement. This implies that there exists no local
measurement that could be performed on the satellite that would in
practice reveal the presence of the superposition. The different
orientations of Hyperion are thus dynamically
environment-superselected
\cite{Zurek:1981:dd,Zurek:1982:tv,Zurek:1993:pu} as the robust
quasiclassical states between which coherence becomes locally
suppressed, thereby ensuring effective classicality for the
satellite. Of course, this conclusion is rather insensitive to the
particular decoherence model: Any environmental monitoring of the
position and orientation of the satellite, such as that mediated by
the ubiquitous scattering of environmental particles
\cite{Joos:1985:iu}, will bring about such decoherence.

Although Zurek's conclusions are intuitively reasonable and in
agreement with general insights gained from studies of environmental
entanglement and decoherence
\cite{Joos:2003:jh,Zurek:2002:ii,Schlosshauer:2007:un}, they were
subsequently criticized by Wiebe and Ballentine
\cite{Wiebe:2005:lm}. These authors revisited the problem of the
quantum--classical correspondence for Hyperion by presenting detailed
numerical studies for an explicit model of the satellite. Based on
their results, they concluded that even in the absence of
environmental interactions there is no problem with the
quantum--classical correspondence for Hyperion. \emph{A forteriori},
this conclusion led the authors to claim that ``decoherence is not
essential to explain the classical behavior of macroscopic bodies''
\cite[p.~022109-1]{Wiebe:2005:lm}, in contrast with Zurek's original
argument \cite{Zurek:1998:om} and the commonly accepted wisdom about
the role of decoherence in the problem of the quantum-to-classical
transition
\cite{Joos:1985:iu,Joos:2003:jh,Zurek:2002:ii,Schlosshauer:2007:un}.

In the following we will not only show that the studies of Zurek, and
of Wiebe and Ballentine, address different problems, but also
demonstrate that the conclusions drawn from the results of these
studies are based on distinct sets of assumptions. By presuming a
strictly epistemic ensemble (statistical) interpretation of quantum
mechanics, Wiebe and Ballentine take the view from the outset that
there is no measurement problem, while this is precisely the problem
addressed by Zurek. In this way, the calculations of Wiebe and
Ballentine do not challenge the conclusions of Zurek's analysis
regarding the role of decoherence, contrary to the claims put forward
in \cite{Wiebe:2005:lm}. In bringing out the fundamental differences
between the two studies, we thus show that the controversy is rooted
in an instance of ``comparing apples and oranges.''  Finally, while
this article is motivated by the specific example of Hyperion and the
corresponding investigations of Zurek, and of Wiebe and Ballentine, it
has a much broader scope: It sheds light on tacit assumptions in the
ensemble interpretation that effectively amount to presuming
characteristic features of classicality usually derived from
decoherence.

\section{What Classicality? An Analysis of Different Paradigms}

In their study \cite{Wiebe:2005:lm}, Wiebe and Ballentine define
proper quantum--classical correspondence for Hyperion via the
condition
\begin{align}
  \label{eq:4}
  \Delta_\text{qm--cl} \equiv \sum_m \abs{ P_\text{cl}(m) -
    P_\text{qm}(m) } \ll 1,
\end{align}
where $P_\text{cl}(m)$ and $P_\text{qm}(m)$ are the classical and
quantum probability distributions, respectively, for Hyperion's
angular momentum along the $z$ axis, which is the space-fixed axis
perpendicular to the orbital plane of the satellite. In the classical
case, the probability distribution is given by the (classically)
incompletely defined initial state evolved under the appropriate
equations of motion. In the quantum setting, Hyperion is described by
a pure-state superposition of eigenstates $\ket{m}$ of the
angular-momentum operator $\op{J}_z$,
\begin{align}
  \label{eq:1}
  \ket{\psi} = \sum_m c_m \ket{m},
\end{align}
and $P_\text{qm}(m)$ is the corresponding probability for finding the
value $m$ in a measurement of the operator-observable $\op{J}_z$,
\begin{align}
  \label{eq:3}
  P_\text{qm}(m) = \abs{ \braket{m}{\psi} }^2 = \abs{c_m}^2.
\end{align}
If the inequality \eqref{eq:4} is fulfilled, i.e., if the classical
and quantum probability distributions $P_\text{cl}(m)$ and
$P_\text{qm}(m)$ agree reasonably well, then Wiebe and Ballentine take
this result as saying that there is no quantum--classical problem for
Hyperion, i.e., that Hyperion behaves classically. Through numerical
studies of an explicit model, the authors find that the inequality
\eqref{eq:4} indeed holds to a sufficient degree.

However, agreement of classical and quantum probability distributions
for a single observable is not a sufficient criterion for proper
quantum--classical correspondence. In the classical setting, both the
value of angular momentum along the $z$ axis and the position
(orientation) of Hyperion are simultaneously well-defined; any
probabilistic aspect is simply due to our (practically motivated, but
not fundamentally required) ignorance about the initial state. In the
quantum setting, on the other hand, the position operator and
$\op{J}_z$ do not commute. This allows for two possible scenarios.

In the first scenario, despite the noncommutativity of these two
operators, we may suppose that the eigenstates of $\op{J}_z$ are also
\emph{approximate} position eigenstates for Hyperion. In this case, a
measurement of $\op{J}_z$ would be unable to distinguish the coherent
superposition of macroscopically distinct positions of Hyperion from
the corresponding classical mixture of positions. Therefore $\op{J}_z$
would simply be the wrong choice of observable for detecting this
nonclassical superposition. However, there always exists a projective
observable that would optimally verify the existence of the
superposition, while \emph{practical} difficulties in measuring such
an observable can in turn be \emph{explained} by environmental
interactions, namely, by environment-induced superselection
\cite{Zurek:1981:dd,Zurek:1982:tv,Zurek:1993:pu}.

In the second scenario, we shall conversely assume that a measurement
of $\op{J}_z$ is indeed sensitive (in the above sense) to the
superposition of macroscopically distinct positions of Hyperion. If we
measure $\op{J}_z$ and obtain a certain outcome, we may thus conclude
that we have measured such a superposition---i.e., that we have
verified the presence of coherence between different positions.  But
Wiebe and Ballentine \emph{a priori} do not regard this thus-confirmed
existence of the superposition as a nonclassical state of affairs.
Instead, Wiebe and Ballentine explicitly state that the only role of
quantum states is to describe ``the probabilities of the various
possible phenomena'' \cite[p.~022109-1]{Wiebe:2005:lm}. Thus they
presume the ensemble, or statistical, interpretation of quantum
mechanics \cite{Ballentine:1970:cz}.

The key assumption of this type of interpretation is to consider the
quantum state as only representing the statistical properties of an
ensemble of similarly prepared systems. The ensemble interpretation
thus implies that the entire formal body of quantum mechanics (for
example, a probability amplitude) has no direct physical meaning, in
the sense of having no direct correspondence to the entities of the
physical world (see also the comment by Leggett
\cite{Leggett:2002:uy}). This interpretation effectively points toward
the need for some hidden-variables theory to fully specify the state of
individual systems, but it does not actually specify what this
``complete theory'' would be.

A signature of the ensemble view is its interpretation of
superpositions. In his review paper \cite{Ballentine:1970:cz},
Ballentine used the example of the momentum eigenstate of a single
electron, which yields a plane wave in configuration space (i.e., a
superposition of all spatial positions). In the ensemble
interpretation, this quantum state is viewed as representing an
(conceptual, infinitely large) ensemble of single electrons with the
same momentum, but evenly spread out over all positions. In other
words, superpositions are interpreted as representing the results of
an ensemble of yet-to-be-performed measurements, while the occurrence
of the individual measurement outcomes is not dynamically explained
(as in explicit hidden-variables or physical-collapse theories) or
represented (as in the collapse postulate) within the
quantum-mechanical formalism. 

Thus, in this interpretation, the superposition considered by Zurek is
simply viewed as describing the probability distribution for finding
the satellite at a particular position and orientation upon
measurement. The measurement problem and the problem of the
quantum-to-classical transition, however, are precisely concerned with
explaining the workings of this instrumentalist algorithm in terms of
a physical theory. Furthermore, and this is the important point, if
Hyperion is considered as a closed system (i.e., if decoherence is
absent) there always exists some observable for Hyperion that would
confirm the presence of the coherent superposition of macroscopically
distinguishable orientations which Zurek has shown to result from the
combination of classical chaotic dynamics and quantum unitary
evolution.

The key question thus is: Why is it so prohibitively difficult to
confirm in practice the presence of coherence in such cat-like
superpositions? This question is not at all addressed by Wiebe and
Ballentine. In the case of Schr\"odinger's cat, the projective
observable directly verifying the presence of the superposition would
be proverbially nonclassical, namely, of the form $ \op{O}_\text{cat}
\propto (\ketbra{\text{alive}}{\text{dead}} \pm
\ketbra{\text{dead}}{\text{alive}})$.  This observable would therefore
have no counterpart in the classical setting, which would exclude any
possibility of comparing the corresponding classical and quantum
probability distributions. Wiebe and Ballentine's particular choice of
$\op{J}_z$ as the observable of interest allows them to avoid such an
obstacle. Although $\op{J}_z$ corresponds (assuming the second
scenario above) to the measurement of a nonclassical superposition of
positions, it still also represents the measurement of a classical
quantity (namely, angular momentum) taking well-defined values in the
classical model of Hyperion. It is only this peculiar feature of the
chosen observable, together with the adoption of an ensemble
interpretation of quantum mechanics, that permits Wiebe and
Ballentine's analysis to proceed.

It is easy to anticipate that once the problem of the
quantum-to-classical transition is reduced to the comparison of
quantum and classical probability distributions for a single
observable (which, as discussed above, is deliberately chosen such as
to make sense also in the classical setting), environmental
interactions and the resulting decoherence processes will play but a
minor role when compared to a framework in which measurability and the
existence of quasiclassical observables is to be \emph{explained} from
within quantum mechanics. Indeed, when Wiebe and Ballentine
investigate the influence of classical noise on the probability
distributions for $\op{J}_z$, aiming to simulate environmental
interactions and decoherence-like processes (see, however,
\cite{Allinger:1995:tq,Joos:2003:jh,Myatt:2000:yy,Schlosshauer:2007:un,Zurek:2002:ii}
on important limitations on simulating decoherence by noise), they
(unsurprisingly) find only a comparably small degree of smoothing of
these distributions.  From this finding they conclude that
environmental interactions---and thus what the authors label
``decoherence''---are insignificant to the problem of classicality and
present this argument as a challenge to Zurek's claims.

But this is a fallacious conclusion.  Decoherence allows one to treat
the superposition locally as an apparent ensemble of quasiclassical
configura\-tions, here presumably coherent states well localized in
both position and angular momentum \cite{Paz:1993:ta}. This leads to
the superselection of the preferred quasiclassical observables that
were simply picked out by Wiebe and Ballentine. (Other practical
obstacles, such as the suitable preparation of measurement apparatuses
and the design of appropriate couplings between system and apparatus,
will naturally also play a role.) The similarity between classical and
quantum probability distributions for a particular observable, as
shown by Wiebe and Ballentine, has simply no bearing on Zurek's
argument.

\section{Conclusions}

Our discussion demonstrates the importance of a careful distinction
between mathematical calculations and their proper (physical)
interpretation. The focus of Zu\-rek's argument is to show how
coherent spreading of the wave packet over macroscopic distances may
become relevant also for macroscopic objects, including celestial
bodies long regarded as prime examples of classical systems, and how
decoherence leads to a local (improper) ensemble of narrow wave
packets describing quasiclassical trajectories in the usual sense of
the emergent-classicality program of decoherence. Wiebe and
Ballentine's analysis, on the other hand, is solely concerned with a
comparison of distributions of measurement outcomes whose
probabilistic aspect has two fundamentally different sources
(classical ``ignorance'' vs.\ quantum ``randomness''). What these
authors have demonstrated is that, \emph{if} we measure the
operator-observable $\op{J}_z$ on Hyperion and \emph{if} we assume the
usual measurement axioms of quantum mechanics, then the resulting
distribution of measurement outcomes will be reasonably close to the
classical distribution. In other words, they have successfully shown
that decoherence does not play a crucial role in restoring proper
quantum--classical correspondence in the Ehrenfest
\cite{Ehrenfest:1927:po} or quantum-Liouville sense for a particular
observable that they deem to be most natural.

But as we have demonstrated in this paper, contrary to the authors'
claims these results do not challenge Zurek's conclusion regarding the
importance of decoherence for the problem of the quantum-to-classical
transition. Zurek's point is that there always exist \emph{some}
observable that could confirm the presence of the nonclassical
superposition state of Hyperion.  Decoherence, then, explains the
practical difficulty in measuring such observables and allows us, for
all practical purposes, to describe measurement locally in effectively
classical terms and thus to ignore the Schr\"odinger-cat problem
\emph{in practice}. The empirical adequacy of Wiebe and Ballentine's
\emph{a priori} belief in the absence of any such problem may thus be
regarded as a \emph{consequence} of the ubiquitous action of
decoherence. We therefore suggest that the authors' conclusion that
``it is not correct to assert that environmental decoherence is the
root cause of the appearance of the classical world''
\cite[p.~022109-13]{Wiebe:2005:lm} is a \emph{non sequitur}.  In this
way, we hope that our analysis has quite peacefully resolved the
controversy.

\begin{acknowledgements}
  The author would like to thank E.\ Joos, H.-D.\ Zeh, W.\ H.\ Zurek,
  and the anonymous referees for valuable comments and
  suggestions. This work was supported by the Australian Research
  Council.
\end{acknowledgements}

\end{document}